\documentclass[twocolumn,nofootinbib,amsmath,amssymb,aps,prd,superscriptaddress,floatfix]{revtex4-2}
\usepackage{amsmath,amssymb,graphicx,xspace,subfigure,tikz,bm}
\usepackage{mathrsfs}

\usepackage[normalem]{ulem}

\usepackage[dvipsnames]{xcolor}
\usepackage{hyperref}
\usepackage{cleveref}
\definecolor{green2}{cmyk}{0.27, 0, 1, 0.52} % MA's nice green
\hypersetup{
    colorlinks=true,       % false: boxed links; true: colored links
    linkcolor=BrickRed,          % color of internal links
    citecolor=green2,        % color of links to bibliography
    filecolor=magenta,      % color of file links
    urlcolor=green2% color of external links
}

\newcommand{\placeholder}[1]{\fbox{\parbox{0.90\linewidth}{\centering #1}}}
\newcommand{\optionalfigure}[2]{\IfFileExists{#1}{\includegraphics[width=\linewidth]{#1}}{\placeholder{#2}}}

\newcommand{\CYTools}[1]
{\texttt{CYTools} }

\newtheorem{lemma}{Lemma}
\newcommand{\su}{\mathfrak{su}}
\newcommand{\so}{\mathfrak{so}}
\newcommand{\uu}{\mathfrak{u}}
\newcommand{\usp}{\mathfrak{usp}}
\newcommand{\cF}{\mathcal{F}}
\newcommand{\Fix}{\operatorname{Fix}}
\newcommand{\Ext}{\operatorname{Ext}}

\begin{document}

% \title{Componentwise D7-tadpole cancellation and the Standard Model:\\
% a no-go for smooth O7$^{-}$ orientifolds}
\title{Can componentwise D7-tadpole cancellation yield a chiral Standard Model?\\
A no-go for smooth O7$^{-}$ orientifolds}
\author{Mudit Jain}
\affiliation{Department of Physics and Astronomy, Rice University, Houston, Texas 77005, U.S.A.}

\date{\today} 

\begin{abstract}
We study a componentwise prescription for D7-tadpole cancellation that includes, as a special case, the local cancellation used in several moduli-stabilisation constructions. We first consider smooth Type IIB Calabi--Yau orientifolds with O7$^{-}$ components of pairwise distinct divisor classes. For each component class $[R_i]$, eight covering-space D7-branes are distributed among smooth irreducible representatives of that class in an orientifold-compatible way. Each component sector may therefore contain several orthogonal, unitary, or symplectic parent factors. We show that these configurations cannot supply the chiral Standard Model through their perturbative D7 spectrum. Without symplectic factors, the entire parent algebra of each component sector embeds into $\mathfrak{so}(8)$ and cannot contain commuting color and weak-isospin subalgebras. Disjointness excludes massless fermions connecting different component sectors, so quark-doublet fermions are absent. Configurations containing symplectic factors require a separate argument: massless hypercharge, together with the same disjointness, forces the net quark-doublet chirality to vanish. Both arguments allow admissible D7 gauge backgrounds and diagonal gauge embeddings, with no restriction on $h^{1,1}_{-}(X)$. We then allow repeated O7 classes, whose D7 sectors can have larger parent groups but necessarily have zero net chirality for both Abelian and non-Abelian gauge backgrounds. We also extend the obstruction to mixed allocations with Whitney remainders supporting no continuous gauge group. We give a geometric condition under which the tadpole forces purely local cancellation. For a specified finite set of allowed D7 support classes, a separate test identifies component sectors in which the tadpole forces componentwise cancellation. We discuss applications to toric constructions and axion potentials.
\end{abstract}

\maketitle

%\tableofcontents

%%%%%%%%%%%%%%%%%%%%%%%%%
\section{Motivation and relation to previous work}
%%%%%%%%%%%%%%%%%%%%%%%%%

Local D7-tadpole cancellation is used in moduli-stabilisation constructions and orientifold surveys. We examine whether a componentwise extension of this prescription can supply a chiral Standard Model through the perturbative D7 spectrum. Each geometric O7 component $R_i$ carries a charge proportional to its divisor class $[R_i]$. We first take these classes to be distinct. We assign the compensating charge $+8[R_i]$ to eight covering-space D7-branes, distributed among smooth representatives of that class together with their orientifold images. All stacks associated with the same class form one \emph{component sector}. Thus \emph{componentwise} refers to the charge allocation associated with each O7 component; cancellation need not hold pointwise on the O7 locus. This charge allocation is an additional restriction beyond the general D7-tadpole condition.

Purely local cancellation places all eight branes of each component sector on $R_i$, giving maximal parent algebra $\bigoplus_{i=1}^{n}\so(8)_i$. More general allocations may place some or all of these branes on other representatives of the same class. If the parent factors are all orthogonal or unitary, the complete algebra of each component sector still embeds into $\so(8)$, and one group-theoretic argument covers all such configurations. Symplectic factors require a separate argument. They can allow quark-doublet fermions, either within one stack or between stacks of the same component class, but massless hypercharge forces their net chirality to vanish.

The obstruction also holds when O7 components share a divisor class. Their combined D7 sector can have a larger parent algebra. Distinct representatives of that class are nevertheless disjoint, and their normal-bundle first Chern class vanishes. These geometric properties exclude net chirality even for non-Abelian gauge backgrounds.

We also consider mixed configurations in which part of each component's compensating charge is carried by fully recombined Whitney branes supporting no continuous gauge group~\cite{Collinucci:2008pf,Crino:2022zjk}. Ordinary stacks arising in their degeneration limits are included, provided their supports satisfy the same smoothness and class conditions as the other ordinary stacks. Section~\ref{subsec:whitney-remainder} shows that the obstruction persists for these mixed configurations.

The purely local case is the standard ``local D7-tadpole cancellation'' prescription. Its relation to a constant axio-dilaton is illustrated by the orientifold construction of Ref.~\cite{Sen:1996vd}. It has appeared as an explicit ingredient in moduli-stabilisation constructions~\cite{Denef:2005mm,Lust:2006zg,Demirtas:2021nlu,McAllister:2024lnt}. It has also been used in studies of instanton corrections~\cite{Blumenhagen:2012kz}, flux vacua~\cite{Coudarchet:2023mmm}, and perturbative corrections~\cite{Gao:2022uop}.

Local cancellation is also adopted in constructions of orientifold geometries~\cite{Lust:2006zh} and used as a reference configuration in geometric surveys \cite{Carta:2020ohw,Crino:2022zjk,Altman:2021pyc}. Several studies compare the local prescription with more general non-local D7 configurations, including Whitney-brane constructions
\cite{Carta:2020ohw,Crino:2022zjk,Shukla:2022dhz,Gao:2022fdi}. General accounts of orientifold model building and its relation to F-theory can be found in Refs.~\cite{Blumenhagen:2006ci,Weigand:2010wm}. An explicit example with several disjoint O7 components, including both locally cancelled and non-invariant D7 configurations, is discussed in Ref.~\cite{Cicoli:2016xae}.

The interplay between chiral D7 sectors and nonperturbative moduli stabilisation was studied in Ref.~\cite{Blumenhagen:2007sm}. Ref.~\cite{Blumenhagen:2012kz} adopts purely local cancellation as a simplifying assumption and notes the need to relax it for concrete model building. We ask whether the broader componentwise prescription can supply the chiral Standard Model through its perturbative D7 spectrum after turning on gauge fluxes. Comparing the parent groups with the Standard Model gauge group does not settle this question. Fluxes can break a parent gauge group and generate chiral matter. As an illustration, Ref.~\cite{Cicoli:2017shd} studies an O7/D7 stack with flux breaking $SO(8)$ to $U(4)$ and producing chiral antisymmetric matter. The obstruction established here concerns the \emph{Standard Model spectrum}, not chirality in general.

A quark doublet must carry both color and weak-isospin charges. Within the perturbative D7 spectrum, these charges must either arise within one component sector or from a string connecting two different sectors. For configurations with distinct O7 classes and no symplectic factors, we exclude both possibilities: each component sector's parent algebra embeds into $\so(8)$ and cannot contain commuting color and weak-isospin algebras, while different component sectors have disjoint supports. The absence of massless matter connecting disjoint D7 sectors was used explicitly in the $SO(8)^{12}$ construction of Ref.~\cite{Denef:2005mm}; another construction involving disjoint smooth O7 components appears in Ref.~\cite{Blumenhagen:2009up}. The required representation-theoretic ingredients are standard; see, for example, Ref.~\cite{Slansky:1981yr}.

We define the componentwise prescription and state the obstruction in Sec.~\ref{sec:statement}. After establishing the geometric and representation-theoretic ingredients, we present both arguments in Sec.~\ref{sec:proof}. For configurations with pairwise distinct O7 classes and no symplectic factors, the obstruction follows from the parent gauge algebras and disjointness of different component sectors. For configurations containing symplectic factors, we show that nonzero net quark-doublet chirality would necessarily make hypercharge massive. The detailed derivation of this second argument is given in Appendix~\ref{app:usp}. Both arguments include split allocations and diagonal gauge embeddings across component sectors. We then extend the zero-index conclusion to repeated O7 classes in Sec.~\ref{subsec:repeated-o7-classes}. We also examine when the tadpole itself fixes the charge allocation. A geometric condition forces purely local cancellation, while a separate test for a finite set of allowed D7 support classes identifies component sectors in which the tadpole forces componentwise cancellation. We then discuss applications to toric hypersurfaces and the consequences of these allocations for axion potentials.

%%%%%%%%%%%%%%%%%%%%%%%%%
\section{Setup and statement}
\label{sec:statement}
%%%%%%%%%%%%%%%%%%%%%%%%%

Let $X$ be a smooth compact Calabi--Yau threefold with a holomorphic involution $\sigma$ satisfying $\sigma^*\Omega_X=-\Omega_X$. We work in the perturbative geometric Type IIB orientifold description. Write the divisorial fixed locus and its class as
\begin{equation}
 R=\bigcup_{i=1}^{n}R_i,
 \qquad [R]=\sum_{i=1}^{n}[R_i],
 \label{eq:R}
\end{equation}
where $R_i$ are the connected components of the divisorial fixed locus, each supporting an O7$^{-}$ plane. Throughout, ``O7 component'' refers to the geometric divisor $R_i$, while $[R_i]$ denotes its cohomology class. These divisors are smooth and mutually disjoint, as explained later. Isolated O3 fixed points, if present, do not enter the argument. We first assume that the O7-component classes are pairwise distinct,
\begin{equation}
 [R_i]\ne[R_j]\qquad(i\ne j).
 \label{eq:distinct-classes}
\end{equation}
We relax this assumption for the zero-index conclusion in Sec.~\ref{subsec:repeated-o7-classes}. We use ``component class'' for $[R_i]$ and ``D7 representative'' for a divisor in that class. 

Our \emph{componentwise D7-tadpole cancellation prescription} is the following. For each $i$, choose smooth irreducible divisors $E_{i\alpha}$ with $[E_{i\alpha}]=[R_i]$.\footnote{For $h^{0,1}(X)=0$ and $k\geq2$, the integral class $\sum_{a=1}^{k}[R_{i_a}]$ of distinct O7 components admits no smooth irreducible holomorphic representative. By the standard divisor exact sequence, every smooth representative of this class has $k$ disjoint connected components, as does the union $\bigcup_{a=1}^{k}R_{i_a}$~\cite{Hartshorne1977,KanazawaWilson2014}.} Write $E'_{i\alpha}=\sigma(E_{i\alpha})$. Count each representative together with its image only once, combining coincident choices into one stack. Place $N_{i\alpha}$ D7-branes on $E_{i\alpha}$ and their images on $E'_{i\alpha}$, where
\begin{equation}
 \sum_\alpha N_{i\alpha}=4,
 \qquad N_{i\alpha}\in\mathbb Z_{>0},
 \label{eq:component-multiplicities}
\end{equation}
with no additional D7-branes. Since $[R_i]$ is orientifold-even,
\begin{equation}
 \begin{gathered}
 [E'_{i\alpha}]=[E_{i\alpha}]=[R_i],\\
 \sum_\alpha N_{i\alpha}
 \bigl([E_{i\alpha}]+[E'_{i\alpha}]\bigr)=8[R_i].
 \end{gathered}
 \label{eq:component-allocation}
\end{equation}
Summing over $i$ satisfies the D7-tadpole condition. All stacks and images assigned to $[R_i]$ form its component sector. The allocation may occupy a single representative and its image, or several such choices within the same class.

The following list describes an individual stack together with its image. Different types may occur within the same component sector.

\begin{enumerate}
 \item \emph{O7-coincident:} $E_{i\alpha}=R_i$. The $2N_{i\alpha}$ covering-space branes lie on the O7$^{-}$ component and have maximal parent algebra $\so(2N_{i\alpha})$.

 \item \emph{Non-invariant:} $E'_{i\alpha}\ne E_{i\alpha}$. There are $N_{i\alpha}$ branes on each of the two image divisors, giving maximal parent algebra $\uu(N_{i\alpha})$.

 \item \emph{Setwise invariant but distinct from the O7 component:} $E'_{i\alpha}=E_{i\alpha}\ne R_i$. The $2N_{i\alpha}$ covering-space branes lie on this invariant divisor. With the symplectic Chan--Paton projection, their maximal parent algebra is $\usp(2N_{i\alpha})$.
\end{enumerate}

There is at most one O7-coincident parent stack in each component sector, since all branes on $R_i$ are grouped together. Our notation $\usp(2N)$ uses the dimension $2N$ of the fundamental representation.

The maximal parent gauge algebra is
\begin{equation}
 \mathfrak g_{\max}
 =\bigoplus_{i=1}^{n}\mathfrak g_i,
 \qquad
 \mathfrak g_i=\bigoplus_\alpha\mathfrak g_{i\alpha},
 \label{eq:parent}
\end{equation}
where $\mathfrak g_{i\alpha}$ is the algebra associated with the corresponding stack. Here \emph{maximal} means before breaking by gauge backgrounds. Each component sector contains eight covering-space D7-branes, regardless of how its allocation is split. An unsplit allocation has $N_{i1}=4$ and gives $\so(8)$, $\uu(4)$, or $\usp(8)$.

Purely local cancellation places all eight branes of every component sector on its O7 component, giving $\bigoplus_i\so(8)_i$. If some branes occupy other representatives, cancellation in that sector holds in cohomology rather than pointwise on $R_i$.

If $R_i$ is the unique effective representative of its class, the eight covering-space D7-branes assigned to $[R_i]$ must lie on $R_i$. If this holds for every $i$, only purely local cancellation is possible, with maximal parent algebra $\bigoplus_{i=1}^{n}\so(8)_i$. Rigidity ensures this uniqueness within a linear-equivalence class.\footnote{For a smooth divisor $R_i$ in a Calabi--Yau threefold, $h^{0,2}(R_i)$ counts infinitesimal geometric deformations. When $h^{0,2}(R_i)=0$, the divisor is rigid and is the unique effective representative in its linear-equivalence class.}

We allow orientifold-compatible holomorphic gauge bundles and worldvolume fluxes satisfying the relevant quantization and tadpole conditions. We also allow Wilson lines and Higgs backgrounds. Throughout, the D7 supports must remain those specified above, and the surviving gauge generators must lie in the parent algebra \eqref{eq:parent}. We use the usual perturbative geometric open-string spectrum. Transitions to singular configurations with additional light sectors lie outside the setup.\\

\noindent{\bf Proposition: Obstruction to a D7 Standard Model:} \textit{In the setup above, suppose that color, weak isospin, and hypercharge are embedded in the unbroken D7 gauge algebra, with a massless hypercharge gauge boson before electroweak symmetry breaking. The net chiral index of the massless D7--D7 quark-doublet fermions is then zero: $\mathcal I_Q=N_Q-N_{\overline Q}=0$, where $Q=(\mathbf3,\mathbf2)_{1/6}$ and $\overline Q=(\overline{\mathbf3},\mathbf2)_{-1/6}$, and the multiplicities count left-handed Weyl multiplets in these representations.\footnote{Equivalently, one may write $\mathcal I_Q=n_L(Q)-n_R(Q)$. An independent right-handed field in $Q$ can be represented by its left-handed charge conjugate in $\overline Q$, so $n_R(Q)=n_L(\overline Q)$ when translating to the all-left-handed convention. This does not count the ordinary antiparticle excitations of a left-handed $Q$ field as an additional left-handed $\overline Q$ field.} The chiral Standard Model, which requires $\mathcal I_Q=3$, is therefore excluded. The zero-index conclusion also holds without~\eqref{eq:distinct-classes}, with all stacks assigned to the same class combined into one component sector. If the O7-component classes are pairwise distinct and no symplectic factors are present, a stronger statement holds: no massless D7--D7 fermion transforms nontrivially under both color and weak isospin, for any hypercharge assignment.}\\

The proposition allows diagonal embeddings across component sectors and imposes no condition on $h^{1,1}_{-}(X)$. Finite central quotients of the gauge group do not change the argument.

Section~\ref{sec:proof} first treats distinct O7 classes, both without and with symplectic factors, including split allocations, and then allows repeated classes. Appendix~\ref{app:usp} gives the detailed derivation of the symplectic argument.

%%%%%%%%%%%%%%%%%%%%%%%%%
\section{Two ingredients}
%%%%%%%%%%%%%%%%%%%%%%%%%

For configurations with pairwise distinct O7 classes and no symplectic factors, the obstruction follows from two facts. First, the distinct O7-component classes have mutually disjoint smooth D7 representatives in the stated setup. Second, the complete parent algebra of each component sector embeds into $\so(8)$, which cannot contain commuting color and weak-isospin algebras. We establish these statements before applying them to the D7 spectrum. The geometric result also enters the hypercharge argument in Sec.~\ref{subsec:symplectic-proof}, whose detailed derivation is given in Appendix~\ref{app:usp}.

\subsection{Disjoint O7 components and their D7 representatives}

Let $p$ be a fixed point of the involution $\sigma$, so that $\sigma(p)=p$. Starting with any Riemannian metric $g$ on $X$, the averaged metric $(g+\sigma^*g)/2$ is invariant under $\sigma$. Thus $\sigma$ is an isometry of this metric and maps geodesics to geodesics.

Write
\begin{equation}
 L_p:=d\sigma_p:T_pX\longrightarrow T_pX
\end{equation}
for the differential, or tangent map, of $\sigma$ at $p$. It describes how $\sigma$ transforms an initial velocity: if $\gamma(0)=p$ and $\gamma'(0)=v$, then $L_pv = d\sigma(\gamma(t))/dt|_{t=0}$. Because $p$ is fixed, both the original and transformed velocities belong to the same tangent space $T_pX$. Moreover, $\sigma^2=\mathrm{id}$ implies $L_p^2=\mathrm{id}$.

The exponential map of the invariant metric therefore satisfies
\begin{equation}
 \sigma\bigl(\exp_p v\bigr)=\exp_p\bigl(L_pv\bigr).
\end{equation}
Here $\exp_p v$ is the point reached at parameter time one by the geodesic starting at $p$ with initial velocity $v$.\footnote{In flat Euclidean space, $\exp_p v=p+v$. On a curved manifold, it is the geodesic version of moving away from $p$ with initial velocity $v$.} The identity says that following this geodesic and then applying $\sigma$ gives the same endpoint as first transforming the initial velocity by $L_p$ and then following the resulting geodesic.

Choose a sufficiently small ball $B$ around the origin in $T_pX$, on which $\exp_p$ is one-to-one. A nearby point $\exp_p v$ is fixed precisely when $L_pv=v$. Consequently,
\begin{equation}
 \Fix(\sigma)\cap\exp_p(B)
 =
 \exp_p\bigl(B\cap\ker(L_p-\mathrm{id})\bigr).
\end{equation}
The space $\ker(L_p-\mathrm{id})$ is the $+1$ eigenspace of $L_p$. It is a linear subspace of $T_pX$, so its image under these local coordinates is a smooth submanifold. Since the involution is holomorphic, its fixed components are also complex submanifolds.

Distinct divisorial branches cannot cross: even if each branch is smooth, their union would fail to be a smooth submanifold at a crossing. The divisorial fixed locus therefore consists of smooth, mutually disjoint components:
\begin{equation}
 R_i\cap R_j=\varnothing\qquad(i\ne j).
 \label{eq:disjoint}
\end{equation}

Let $E_i$ and $E_j$ denote arbitrary individual D7 supports from component sectors $i$ and $j$, including any image supports. Then $[E_i]=[R_i]$ and $[E_j]=[R_j]$. For $i\ne j$ and a K\"ahler form $J$,
\begin{equation}
 \int_X[E_i]\wedge[E_j]\wedge J
 =\int_X[R_i]\wedge[R_j]\wedge J=0.
 \label{eq:representatives}
\end{equation}
By \eqref{eq:distinct-classes}, $E_i$ and $E_j$ are distinct. If they intersected, their intersection would consist of holomorphic curves counted with positive multiplicities. The integral of the K\"ahler form $J$ over each such curve is its positive area. Their intersection would therefore give a positive value in \eqref{eq:representatives}, contradicting its vanishing. Thus every support in one component sector is disjoint from every support in another. Representatives within the same component class may intersect, including brane--image pairs and separately chosen stacks. The absence of \emph{massless} fermions connecting different component sectors follows from the holomorphic description below.

In the holomorphic large-volume description, Ext groups describe the relevant open-string fermion zero modes \cite{Katz:2002gh}. Let $\iota_i:E_i\hookrightarrow X$ be the inclusion of a D7 support, and let $\mathcal V_i$ be its gauge bundle. For supports in different component sectors, disjointness gives
\begin{equation}
 \Ext_X^k\!\left(
   \iota_{i*}\mathcal V_i,\iota_{j*}\mathcal V_j
 \right)=0
 \qquad\text{for all }k,\quad i\ne j.
 \label{eq:ext}
\end{equation}
The same statement applies to the brane images.

Here $\iota_{i*}\mathcal V_i$ means the bundle data viewed as a sheaf on $X$, supported on $E_i$ and zero away from it. Because the supports are disjoint, every point of $X$ has a neighborhood in which at least one of the two sheaves is zero. The local Ext sheaves therefore vanish, and so do the global Ext groups.

Vanishing in every degree excludes the entire massless fermion sector connecting the two supports, including vectorlike pairs. This holds for any holomorphic gauge bundles on the disjoint supports.\footnote{This statement concerns fermions and does not assume a common unbroken supersymmetry. Supersymmetry-breaking worldvolume flux can make a stretched-string scalar massless at nonzero brane separation while the connecting fermions remain massive \cite{Hebecker:2011hk}. If the branes preserve a common supersymmetry in a four-dimensional Minkowski background, scalar and fermion superpartners have equal masses, and the connecting scalars are also absent from the massless spectrum.}

%%%%%%%%%%%%%%%%%%%%%%%%%
%%%%%%%%%%%%%%%%%%%%%%%%%
%MJ: Some facts to state maybe?
% A vector bundle \(\mathcal V_i\) assigns an internal vector space to every point of \(D_i\), with consistent rules for relating the spaces in overlapping coordinate patches. Gauge fields are connections on such bundles.
% A sheaf keeps track of data defined on small open sets and the rules for restricting and gluing those data. Its support is where its local data are nonzero.
% The spaces \(\operatorname{Ext}^k\) are mathematical spaces used to describe open-string sectors between these brane objects. The superscript \(k\) labels their degree. Degree zero is a space of maps; higher degrees encode more general extension data. In the supersymmetric geometric description, these graded spaces enter the counting of open-string states. One should not identify every degree with an independent four-dimensional particle species.
%%%%%%%%%%%%%%%%%%%%%%%%%
%%%%%%%%%%%%%%%%%%%%%%%%%

\subsection{Color and weak isospin without symplectic factors}

Consider a component sector containing only orthogonal and unitary factors. A unitary transformation on $N$ complex components acts on $2N$ real components and preserves their Euclidean norm. Its real determinant is $|\det U|^2=1$, so
\begin{equation}
 \uu(N)\subset\so(2N).
 \label{eq:unitary-orthogonal}
\end{equation}
Applying this inclusion to each unitary factor gives
\begin{equation}
 \mathfrak g_i
 \subseteq
 \bigoplus_\alpha\so(2N_{i\alpha})
 \subseteq\so(8).
 \label{eq:sector-so8}
\end{equation}
The middle algebra acts block diagonally on real vector spaces of dimensions $2N_{i\alpha}$. Together these spaces have dimension $\sum_\alpha 2N_{i\alpha}=8$, which gives the last inclusion. This argument applies even when the branes occupy different representatives of the same class. It remains to establish the following fact.

\begin{lemma}
\label{lem:algebra}
The algebra $\so(8)$ has no subalgebra isomorphic to
$\su(3)\oplus\su(2)$.
\end{lemma}

Suppose that $\su(3)$ is embedded in $\so(8)$. By definition, $\so(8)$ acts faithfully on $\mathbb R^8$ by real antisymmetric matrices.\footnote{A Lie-algebra representation is faithful if no nonzero generator acts as the zero matrix.} Any embedded $\su(3)$ therefore acts on this same real eight-dimensional space. We classify the possible color actions and determine which infinitesimal rotations commute with them.

For this purpose, complexify the representation: allow the eight vector coordinates to be complex while retaining the same action matrices. The resulting space $V_{\mathbb C}$ has complex dimension eight. The complex irreducible representations of $\su(3)$ of complex dimension at most eight are $\mathbf1,\mathbf3,\overline{\mathbf3}, \mathbf6,\overline{\mathbf6},\mathbf8$. The representations $\mathbf3$ and $\mathbf6$ are inequivalent to their complex conjugates. Since the original action is real, each must appear in $V_{\mathbb C}$ with the same multiplicity as its conjugate. In particular, a $\mathbf6$ would require an additional $\overline{\mathbf6}$, exceeding the available dimension.

The only possible nontrivial decompositions are therefore
\begin{equation}
 V_{\mathbb C}\big|_{\su(3)}
 =
 \mathbf3\oplus\overline{\mathbf3}\oplus\mathbf1\oplus\mathbf1,
 \qquad\text{or}\qquad
 V_{\mathbb C}\big|_{\su(3)}=\mathbf8.
 \label{eq:branching}
\end{equation}
The all-trivial action is excluded because the defining representation of $\so(8)$ is faithful.

In the first case, the color triplet has three complex components, or equivalently six real components. The remaining two real directions are color singlets. On the triplet, the continuous orthogonal transformations that commute with color are the overall phase rotations, with Lie algebra $\uu(1)$. The two singlet directions admit an independent ordinary rotation, with Lie algebra $\so(2)\simeq\uu(1)$. A transformation commuting with color cannot mix the triplet with the singlets, since they have different color transformation laws. The generators commuting with color therefore form the following algebra, called its centralizer:\footnote{The notation $\mathfrak z_{\so(8)}(\su(3))$ denotes all $A\in\so(8)$ satisfying $[A,C]=0$ for every color generator $C$. Its elements need not commute with one another. Any weak-isospin generators commuting with color would have to belong to this algebra.}
\begin{equation}
 \mathfrak z_{\so(8)}\bigl(\su(3)\bigr)
 =\uu(1)\oplus\uu(1).
\end{equation}

In the second case, the real eight-dimensional representation is the adjoint of compact $\su(3)$. Concretely, its vectors can be represented by traceless Hermitian $3\times3$ matrices $H$, with color acting as $H\mapsto UHU^\dagger$. There are eight real components, and the action preserves the positive definite inner product $\operatorname{tr}(HK)$.

The complexification of this representation is irreducible. Schur's lemma states that an operator commuting with all generator matrices on an irreducible complex representation must be proportional to the identity. A commuting real operator is therefore $a\mathbf1_8$ with $a\in\mathbb R$. To belong to $\so(8)$ it must also be antisymmetric, which requires $a=0$. Thus no nonzero generator in $\so(8)$ commutes with this color action.

In the first case the centralizer is abelian, while in the second it vanishes. Neither can contain $\su(2)$. Since the two factors in $\su(3)\oplus\su(2)$ must commute, the claimed subalgebra cannot exist in $\so(8)$.

Together with \eqref{eq:sector-so8}, Lemma~\ref{lem:algebra} excludes commuting color and weak-isospin algebras from every component sector without symplectic factors, including arbitrary splittings of its charge allocation.

%%%%%%%%%%%%%%%%%%%%%%%%%
\section{Proof of the obstruction}
\label{sec:proof}
%%%%%%%%%%%%%%%%%%%%%%%%%

We first treat pairwise distinct O7 classes. The two ingredients established above exclude massless quark-doublet fermions when no symplectic factors are present. With symplectic factors, requiring massless hypercharge forces their net chirality to vanish. We then allow repeated O7 classes and establish zero net chirality in those sectors directly from their geometry.

\subsection{Without symplectic factors}
\label{sec:no-symplectic-proof}

Consider any embedding of $\su(3)_C\oplus\su(2)_L$ into $\mathfrak g_{\max}=\bigoplus_i\mathfrak g_i$. The embedding may be diagonal: the same color or weak gauge transformation can act simultaneously on several component sectors.

For each component sector, examine the projected color and weak generators in its complete parent algebra $\mathfrak g_i$. Because $\su(3)$ and $\su(2)$ are simple Lie algebras, each projected action is either zero or faithful: a nonzero projection represents the entire corresponding algebra without losing any independent generator.\footnote{The generators mapped to zero remain mapped to zero after taking a commutator with any generator. For a simple Lie algebra, such a subspace is either zero or the whole algebra.} If both actions were nonzero, their commuting images would give $\su(3)\oplus\su(2)$ inside $\mathfrak g_i$ and hence inside $\so(8)$ by \eqref{eq:sector-so8}, contradicting Lemma~\ref{lem:algebra}. Therefore at least one of color and weak isospin acts trivially on the entire component sector.

Equation~\eqref{eq:ext} excludes massless fermions connecting different component sectors. Every massless D7--D7 fermion therefore receives its gauge charges from a single sector, although its endpoints may belong to different stacks within that sector. This includes strings between a representative and its image, as well as strings between separately chosen representatives of the same class. Since at least one of color and weak isospin acts trivially throughout the sector, no such fermion can transform nontrivially under both.

Admissible Abelian or non-Abelian gauge fluxes, Wilson lines, and Higgs backgrounds can change the unbroken gauge algebra and matter spectrum. They cannot enlarge the parent algebra or create a massless fermionic sector between disjoint supports. The proof therefore remains valid whenever the supports and parent algebras remain those specified in Sec.~\ref{sec:statement}; no classification of gauge backgrounds is needed.

\subsection{Symplectic factors and massless hypercharge}
\label{subsec:symplectic-proof}

When symplectic factors are present, color and weak isospin can act nontrivially within the same component sector. Quark doublets may then arise within one stack or between different stacks of that class. The obstruction instead comes from requiring massless hypercharge: the flux needed for nonzero net quark-doublet chirality necessarily gives hypercharge a St\"uckelberg mass. We explain this connection here and give its detailed derivation in Appendix~\ref{app:usp}.

Consider a component sector $i$ capable of supplying an SM quark doublet. It has eight covering-space Chan--Paton labels, which are the gauge labels carried by the string endpoints. The color triplet must occur together with its conjugate, using six of these labels. The remaining two form the weak doublet, which must lie in a symplectic parent factor. Appendix~\ref{app:usp} establishes that the endpoint representation has the form
\begin{equation}
 \begin{aligned}
 V_i={}&(\mathbf3,\mathbf1)_{q_i} \oplus(\overline{\mathbf3},\mathbf1)_{-q_i} \oplus(\mathbf1,\mathbf2)_0.
 \end{aligned}
 \label{eq:main-usp-branching}
\end{equation}
The color and image-color blocks have opposite hypercharges. On the weak doublet, an overall phase would fail to preserve the symplectic pairing, so the commuting hypercharge generator vanishes. All eight endpoint components have been used; there is no additional block in this sector that could contribute independently to the hypercharge-masslessness condition.

The charges in \eqref{eq:main-usp-branching} belong to individual endpoints. A physical D7--D7 field carries two endpoint indices, and its hypercharge is the difference of their charges. A string joining the color and weak blocks can therefore carry $(\mathbf3,\mathbf2)_{q_i}$ even though the weak endpoint has zero hypercharge. For an SM quark doublet, $q_i=1/6$.

Preserving color requires the gauge background to act identically on the three color components. The weak flux is orientifold odd, so its integral against the even restricted class $[R_i]$ vanishes. Since the color and weak supports both have class $[R_i]$, the orientifold-projected net quark-doublet index depends on a single color-flux integral:
\begin{equation}
 \begin{aligned}
 A_{c,i}
 &:=\int_{C_i}[R_i]\big|_{C_i}\wedge f_{c,i},\\
 \mathcal I_{Q_i}&=\epsilon_i A_{c,i}.
 \end{aligned}
 \label{eq:main-usp-Q-index}
\end{equation}
Here $C_i$ supports the color block, $f_{c,i}$ is its dimensionless gauge-invariant flux, including the pulled-back $B$ field, and $\epsilon_i=\pm1$ sets the index convention. Appendix~\ref{app:usp} derives this relation for both coincident and distinct supports, including the projection factor. The fluxes need not be restrictions of bulk flux classes.

The same integral determines the gauging of the $C_4$ axions under hypercharge. In the scalar description, the D7 Chern--Simons couplings make these axions shift under a hypercharge gauge transformation. For an even divisor class $\omega$, the corresponding axion charge is proportional to
\begin{equation}
 K_Y(\omega)
 =
 \sum_S\int_S
 \omega\big|_S\wedge
 \operatorname{tr}(Y_S\cF_S).
 \label{eq:main-usp-K}
\end{equation}
Here the sum runs over distinct occupied D7 support divisors $S$ in the covering space, including orientifold images. All branes sharing a support are included in its Chan--Paton trace. On each support, $Y_S$ is the action of the global hypercharge generator and $\cF_S$ is the internal gauge-invariant field strength. The gauged axion kinetic terms give hypercharge a mass whenever any of these charges is nonzero, because the axion kinetic matrix is positive definite. Massless hypercharge therefore requires
\begin{equation}
 K_Y(\omega)=0
 \qquad
 \text{for every }\omega\in H^{1,1}_{+}(X).
 \label{eq:main-usp-massless-Y}
\end{equation}

Let $C_i'=\sigma(C_i)$ support the image-color block. Its flux satisfies $f_{c',i}=-\sigma^*f_{c,i}$. The color and image-color blocks thus have opposite hypercharges and opposite flux integrals against an even class. Their contributions to $K_Y$ add rather than cancel. When tested against $[R_i]$, each block contributes $3q_i A_{c,i}$, where the factor three counts its color components. The weak block contributes nothing because its hypercharge is zero. The contribution of this component sector is consequently $K_{Y,i}([R_i])=6q_i A_{c,i}$.

Now apply \eqref{eq:main-usp-massless-Y} to $\omega=[R_i]$. Every D7 support $S$ in another component sector is disjoint from $R_i$, so $[R_i]|_S=0$ in $H^2(S,\mathbb R)$. All other component sectors therefore drop out of this particular test. Using \eqref{eq:main-usp-Q-index}, we obtain
\begin{equation}
 \begin{aligned}
 0=K_Y([R_i])
 &=K_{Y,i}([R_i])\\
 &=6q_i A_{c,i}
 =6q_i\epsilon_i\,\mathcal I_{Q_i}.
 \end{aligned}
 \label{eq:main-usp-local-obstruction}
\end{equation}
Since an SM quark doublet has $q_i\ne0$, its sector must have $\mathcal I_{Q_i}=0$. Nonzero net quark-doublet chirality would necessarily make hypercharge massive. Mixing hypercharge with generators in other component sectors cannot cancel this contribution, because their contributions to $K_Y([R_i])$ vanish separately. The argument uses disjointness and does not require linear independence of the component classes.

By \eqref{eq:ext}, every massless D7--D7 fermion belongs to a single component sector. Sectors without symplectic factors cannot supply quark doublets, while every sector that can supply SM quark doublets has zero net index by
\eqref{eq:main-usp-local-obstruction}. Hence
\begin{equation}
 \mathcal I_Q=\sum_i\mathcal I_{Q_i}=0.
 \label{eq:main-zero-Q-index}
\end{equation}
This proves the proposition for pairwise distinct O7 classes, including split and unsplit allocations, mixed products, and diagonal embeddings. Massless vectorlike quark-doublet pairs are allowed, but the net chirality required by the Standard Model is excluded.

\subsection{Repeated O7 classes}
\label{subsec:repeated-o7-classes}

We now allow distinct O7 components to have the same class, while retaining the prescribed smooth irreducible D7 supports. Their assigned branes form one component sector, containing eight covering-space branes for each O7 component in that class. Its combined parent algebra can therefore be larger than $\so(8)$, and the preceding eight-brane argument no longer applies to this sector.\footnote{Branes on the same support form one stack, even if their charges were assigned to different O7 components. A support coinciding with any O7$^{-}$ component has the orthogonal projection.} We show directly from the geometry that this sector has zero net chirality.

Suppose that two distinct O7 components satisfy $[R_i]=[R_j]=[R_{\rm com}]$. Since they are disjoint,
\begin{equation}
 [R_{\rm com}]^2=[R_i]\wedge[R_j]=0.
 \label{eq:repeated-class-square}
\end{equation}
Any two distinct smooth irreducible representatives $S,T$ of $[R_{\rm com}]$ are consequently disjoint. Their intersection would consist of holomorphic curves with positive K\"ahler area, whereas $\int_X[S]\wedge[T]\wedge J=\int_X[R_{\rm com}]^2\wedge J=0$. Thus there are no massless D7--D7 fermions connecting distinct supports within this class sector.

We must also consider modes on a coincident stack with support $S$. Choose an O7 component among $R_i,R_j$ that is distinct from $S$. It is a representative of $[R_{\rm com}]$ disjoint from $S$, so the restriction of $[R_{\rm com}]$ to $S$ vanishes. Therefore
\begin{equation}
 c_1(N_{S/X})=[R_{\rm com}]|_S=0.
 \label{eq:repeated-class-normal}
\end{equation}
For modes on this common support, the net chiral index has the form~\cite{Collinucci:2008pf}
\begin{equation}
 \mathcal I_{AB}
 =\pm\int_S c_1(N_{S/X})\wedge\alpha_{AB}
 =0,
 \label{eq:repeated-class-zero-index}
\end{equation}
where $\alpha_{AB}$ is the two-form class determined by the worldvolume gauge background in that matter sector.
%\footnote{For gauge-bundle blocks $\mathcal V_A,\mathcal V_B$ of ranks $k_A,k_B$, the class is $\alpha_{AB}=k_Ac_1(\mathcal V_B)-k_Bc_1(\mathcal V_A)$, with the overall sign fixed by the index convention. For line bundles this reduces to the corresponding flux difference. No decomposition of a non-Abelian bundle into line bundles is assumed.} 
The vanishing follows from the normal-bundle class, so it holds for Abelian and non-Abelian gauge backgrounds, including fluxes that are not inherited from $X$.

The additional O7 terms in the orientifold-projected indices also vanish. Every O7 component distinct from $S$ is disjoint from it by the same intersection-area argument. If $S$ itself is an O7 component, its self-restriction vanishes by~\eqref{eq:repeated-class-normal}. Hence the total O7 class satisfies $[R]|_S=0$, and its pairing with any worldvolume flux class is zero. The repeated-class sector therefore has zero net chirality for every allowed allocation, independently of whether hypercharge is massless. Larger parent groups may allow quark-doublet representations and vectorlike pairs, but they cannot supply the required net chirality.

After combining equal classes, supports in different component sectors remain disjoint by the same positivity argument. Sectors whose class occurs once contain eight covering-space branes and obey the preceding arguments. Since sectors in repeated classes also have zero net chirality, the total quark-doublet index vanishes even when the O7-component classes are not pairwise distinct.

%%%%%%%%%%%%%%%%%%%%%%%%%
\section{Scope and consequences}
%%%%%%%%%%%%%%%%%%%%%%%%%

For configurations with pairwise distinct O7 classes and no symplectic factors, the obstruction is already the absence of even one massless quark-doublet fermion, independently of hypercharge. Configurations containing symplectic factors can allow such fermions, but massless hypercharge forces their net chirality to vanish. Repeated classes can also allow larger parent groups and vectorlike quark-doublet pairs, but their net indices vanish by the geometric argument of Sec.~\ref{subsec:repeated-o7-classes}. The Standard Model instead requires three net quark doublets. These arguments obstruct the required spectrum before any analysis of realistic Yukawa couplings. A product of parent factors may contain the Standard Model gauge algebra, and fluxed stacks may support chiral matter, without supplying the chiral Standard Model spectrum.

Euclidean D3-brane (ED3) fluxes affect instanton actions and zero modes~\cite{Blumenhagen:2009qh}. An ED3 intersecting several D7 sectors may carry charged zero modes for each, but these are integrated over in the instanton amplitude and are not propagating four-dimensional matter. Such instantons can generate operators involving existing fields; they do not supply new perturbative D7 fermions. Mass terms preserving the Standard Model gauge symmetry remove conjugate pairs and cannot change the net quark-doublet index. No assumption that ED3 contributions vanish is needed.

None of these arguments requires any restriction on $h^{1,1}_{-}(X)$. Smoothness gives disjoint O7 components regardless of the number of orientifold-odd classes. The hypercharge argument tests masslessness against the even class of an individual O7 component, so additional odd classes do not relax the resulting constraint. Equality of divisor classes does not imply equality of supports: distinct brane--image divisors can represent the same class even when $h^{1,1}_{-}(X)=0$.

The prescription assigns eight covering-space D7-branes to each geometric O7 component, including their orientifold images. Their supports must be smooth irreducible representatives of the corresponding component class. When classes repeat, their assigned branes are combined into one component sector as in Sec.~\ref{subsec:repeated-o7-classes}. All distributions satisfying~\eqref{eq:component-multiplicities} are included. Section~\ref{subsec:whitney-remainder} extends the obstruction to mixed configurations with Whitney remainders, including their ordinary D7 limits, while retaining the prescribed support conditions for all ordinary stacks. More general allocations of the gauge stacks and nonperturbative realizations of the visible sector require a separate analysis.

\subsection{Mixed configurations with Whitney branes}
\label{subsec:whitney-remainder}

The compensating charge associated with an O7 component can also be shared between ordinary D7 stacks and Whitney branes. We retain the ordinary supports of Sec.~\ref{sec:statement}: each is a smooth irreducible representative of an individual class $[R_i]$. We now allow these stacks to carry less than $8[R_i]$, with the remaining D7 charge carried by fully recombined Whitney branes supporting no continuous gauge group~\cite{Collinucci:2008pf,Crino:2022zjk}. The characteristic singularities of their worldvolumes are allowed; the Calabi--Yau threefold $X$ and the ordinary D7 supports remain smooth.

Let $\eta_i$ denote the ordinary covering-space D7 charge assigned to the geometric component $R_i$, including images. Each brane--image pair contributes two, so the mixed allocation takes the form
\begin{equation}
 \begin{gathered}
 \eta_i=2\sum_\alpha N_{i\alpha}
 \in\{0,2,4,6,8\},\\
 [D_{\mathrm{ord}}]=\sum_i\eta_i[R_i],
 \qquad
 [D_{\mathrm W}]=\sum_i(8-\eta_i)[R_i].
 \end{gathered}
 \label{eq:mixed-whitney-allocation}
\end{equation}
Here $[D_{\mathrm W}]$ is the total covering-space D7 charge of the Whitney remainder. It may contain several Whitney branes; no separate Whitney support is required for each O7 component. We consider allocations for which such an effective remainder exists. When O7 classes repeat, their ordinary allocations form one component sector, whose total multiplicity may exceed eight.

Ordinary stacks arising when a Whitney configuration factorizes or develops coincident branes must also be included in $\eta_i$. Consider, for example, a class occurring once, with six ordinary covering-space branes and a Whitney remainder of charge $2[R_i]$. If this remainder becomes a brane--image pair on smooth representatives of $[R_i]$, it gives an ordinary $U(1)$ stack. If it becomes two coincident branes on a smooth invariant representative with the symplectic projection, it gives $USp(2)$. In either case these two branes join the ordinary allocation, so the resulting configuration has $\eta_i=8$. These limits are therefore included in the original prescription, provided all ordinary supports satisfy its smoothness and class conditions.

To establish the obstruction for the mixed allocation, first recall that ordinary supports in different component sectors remain disjoint: their classes still satisfy \eqref{eq:representatives}. A quark doublet with both endpoints on ordinary stacks must consequently arise within one component sector. First consider a class occurring once. If $\eta_i<8$, that sector has too few Chan--Paton labels to support both color and weak isospin. The color triplet and its conjugate require at least six labels, while a commuting weak doublet requires at least two more, as in Sec.~\ref{subsec:symplectic-proof}. Strings joining the ordinary stacks to a Whitney brane may still exist, but their Whitney endpoint supplies no color or weak-isospin charge. They cannot supply the missing quark doublets.

It remains to consider a sector with $\eta_i=8$. Its ordinary branes already carry the full charge $8[R_i]$. For any of its supports $E_{i\alpha}$,
\begin{equation}
 \int_X[D_{\mathrm W}]\wedge[E_{i\alpha}]\wedge J
 =
 \sum_j(8-\eta_j)
 \int_X[R_j]\wedge[R_i]\wedge J
 =0.
 \label{eq:whitney-disjoint-full-sector}
\end{equation}
The $j=i$ term vanishes because $\eta_i=8$, and the remaining terms vanish because distinct O7 components are disjoint. All ordinary divisor factors have already been included in $D_{\mathrm{ord}}$, so the Whitney remainder has no divisor component in common with $E_{i\alpha}$. Any nonempty intersection would therefore contain holomorphic curves with positive K\"ahler area, contradicting~\eqref{eq:whitney-disjoint-full-sector}. Thus the Whitney remainder is disjoint from every ordinary support in this sector, including its images.

Any quark doublet arising from a class occurring once is therefore confined to an ordinary component sector with $\eta_i=8$, whose supports and multiplicities are exactly those of the original prescription. Without symplectic factors, the representation-theoretic argument excludes such fermions. With symplectic factors, the same hypercharge-masslessness condition~\eqref{eq:main-usp-local-obstruction} forces their net chirality to vanish. In particular, this includes the $3+1$ split in which the additional $USp(2)$ supplies weak isospin.

For a repeated class $[R_{\rm com}]$, the ordinary supports have zero net chirality by Sec.~\ref{subsec:repeated-o7-classes}. For any such support $S$, its intersection with the Whitney remainder also has zero K\"ahler area, since
\begin{equation}
 \int_X[D_{\mathrm W}]\wedge[S]\wedge J
 =\sum_\ell(8-\eta_\ell)
   \int_X[R_\ell]\wedge[R_{\rm com}]\wedge J
 =0.
 \label{eq:whitney-repeated-class-disjointness}
\end{equation}
Terms from the repeated class vanish because $[R_{\rm com}]^2=0$, and all other terms vanish by disjointness of the O7 components. Since the ordinary divisor factors have already been extracted, positivity of intersection areas makes the Whitney remainder disjoint from these supports. The obstruction therefore also holds for mixed allocations with repeated O7 classes.

\subsection{A geometric condition forcing local cancellation}
\label{subsec:forced-local-cancellation}

A sufficient geometric condition forces purely local cancellation without imposing the componentwise prescription. Suppose there exists a K\"ahler form $J$ such that
\begin{equation}
 \int_X[R_i]^2\wedge J<0
 \qquad\text{for every }i.
 \label{eq:forced-local-cancellation}
\end{equation}

For any D7 configuration with effective holomorphic supports, let $m_i\in\mathbb Z_{\geq0}$ denote the covering-space D7 multiplicity on the actual O7 component $R_i$. Let $S_a$ denote the remaining irreducible support components, with covering-space multiplicities $n_a\in\mathbb Z_{>0}$. Thus $S_a\ne R_i$ for every $a$ and $i$. Define their total charge class by
\[
 [E]:=\sum_a n_a[S_a].
\]
The total covering-space D7 charge is therefore
\[
 [D_{\mathrm{D7}}]=\sum_i m_i[R_i]+[E].
\]

Pairing the general tadpole condition $[D_{\mathrm{D7}}]=8\sum_i[R_i]$ with $[R_i]\wedge J$, and using disjointness of the O7 components, gives
\[
 (8-m_i)\int_X[R_i]^2\wedge J
 =\int_X[E]\wedge[R_i]\wedge J\geq0.
\]
Indeed, each $S_a$ and $R_i$ have no common divisor component. Their intersection consists of holomorphic curves with positive multiplicities, or is empty. Every nonempty such curve has positive area with respect to the K\"ahler form $J$, so the integral on the right is nonnegative.\footnote{If an O7 component $R_i$ is disjoint from every remaining D7 support $S_a$, the same tadpole equation gives $(8-m_i)[R_i]^2=0$ in $H^4(X,\mathbb R)$. Thus $[R_i]^2\ne0$ already forces $m_i=8$, without requiring $\int_X[R_i]^2\wedge J<0$. This conclusion applies to that component independently of the self-intersections of the other O7 components.} The negativity in Eq.~\eqref{eq:forced-local-cancellation} therefore forces $m_i\geq8$ for every $i$.\footnote{The negativity condition also makes $R_i$ rigid: a deformation to a distinct divisor $S$ in the same class would give a nonnegative intersection area, $\int_{R_i\cap S}J=\int_X[R_i]^2\wedge J\geq0$, contradicting the condition.}
%\footnote{The negativity condition also implies that $R_i$ is geometrically rigid: its normal bundle $N_{R_i/X}\simeq\mathcal O_X(R_i)|_{R_i}$ has negative degree with respect to $J|_{R_i}$ and hence admits no nonzero holomorphic sections.}

Pairing the remaining tadpole $\sum_i(m_i-8)[R_i]+[E]=0$ with $J^2$ then forces $m_i=8$ for every $i$ and the remaining sum over $S_a$ to be empty: all coefficients are nonnegative, and every nonzero effective support has strictly positive K\"ahler volume. Thus the tadpole itself requires eight covering-space D7-branes on each $R_i$. The maximal parent algebra is $\bigoplus_i\so(8)_i$, and the orthogonal-case obstruction applies.

Only the existence of one such K\"ahler form $J$ is needed. Any proposed configuration at another K\"ahler point obeys the same cohomological tadpole and can be tested with $J$. The conclusion therefore holds throughout the K\"ahler cone of the fixed complex manifold $X$.

We will apply this criterion to explicit orientifold geometries in a future work and identify the geometries it excludes from admitting a perturbative D7 Standard Model.

\subsection{D7 allocations forced by the tadpole}
\label{subsec:restricted-toric-classes}

The preceding geometric criterion forces purely local cancellation. We now ask what the tadpole forces when the allowed D7 support classes belong to a specified finite set. Let $[D_\alpha]$, with $\alpha=1,\ldots,m$, denote the distinct allowed classes, including the classes of all orientifold images. Each class must admit a smooth irreducible holomorphic representative, but need not be a prime toric class. Membership in the effective cone alone does not guarantee the required representatives~\cite{Gendler:2026uux}. All D7 configurations considered in this test use these classes.

Choose a real basis $[H_A]$ of divisor classes, with $A=1,\ldots,h^{1,1}(X)$, consisting of orientifold-even and orientifold-odd elements. Write
\begin{equation}
 \begin{aligned}
 [D_\alpha]
 &=\sum_{A=1}^{h^{1,1}(X)}d_\alpha^{\,A}[H_A],\\
 [R_i]
 &=\sum_{A=1}^{h^{1,1}(X)}\rho_i^{\,A}[H_A].
 \end{aligned}
 \label{eq:toric-class-expansions}
\end{equation}
Here $d_\alpha^{\,A}$ and $\rho_i^{\,A}$ are the coefficients of the classes in this basis.

Let $x_\alpha$ count covering-space D7-branes in class $[D_\alpha]$, including images and all occupied representatives. The tadpole gives one equation for each basis direction:
\begin{equation}
 \begin{aligned}
 \sum_{\alpha=1}^{m}x_\alpha d_\alpha^{\,A}
 &=8\sum_{i=1}^{n}\rho_i^{\,A},
 &\quad A&=1,\ldots,h^{1,1}(X),\\
 x_\alpha&\geq0,
 &\quad \alpha&=1,\ldots,m.
 \end{aligned}
 \label{eq:restricted-class-tadpole}
\end{equation}
The sum over $i$ includes every O7 component, so repeated O7 classes contribute with their component multiplicities. Since every $[R_i]$ is orientifold even, $\rho_i^{\,A}=0$ in the odd basis directions. The corresponding equations therefore require zero total odd D7 charge.

To determine which allocations the tadpole forces, initially allow real nonnegative brane multiplicities $x_\alpha$. If \eqref{eq:restricted-class-tadpole} has no real solution, the allowed classes cannot cancel the tadpole. Otherwise, pairing the tadpole with $J^2$ fixes the total D7 volume. Since every allowed class has positive K\"ahler volume, each $x_\alpha$ is bounded above. For each class,
find
\begin{equation}
 x_\alpha^{\min}=\min x_\alpha,
 \qquad
 x_\alpha^{\max}=\max x_\alpha,
 \label{eq:restricted-charge-bounds}
\end{equation}
over all real vectors $\mathbf x$ satisfying \eqref{eq:restricted-class-tadpole}.

A zero minimum means that a real allocation can omit that class. A positive minimum means that some charge in the class is compulsory. Equal minimum and maximum values mean that its multiplicity is fixed. Physical allocations must additionally obey integer multiplicities and brane--image counting. If the real bounds already fix a coefficient to eight, every allowed physical allocation also fixes it to eight.

Consider an O7 component whose class occurs among the allowed D7 classes and differs from every other O7-component class. Let $\alpha(i)$ be the index for which $[D_{\alpha(i)}]=[R_i]$. Let $\mathcal F$ contain all such indices $i$ for which
\begin{equation}
 x_{\alpha(i)}^{\min}
 =x_{\alpha(i)}^{\max}=8.
 \label{eq:individually-forced-sector}
\end{equation}
Thus every allowed physical allocation contains exactly eight covering-space D7-branes in class $[R_i]$ for each $i\in\mathcal F$.

For any such configuration, let $[T_a]$ label the remaining occupied D7 classes, with equal classes combined, and let $P_a$ be their positive integer multiplicities. Subtracting $8[R_i]$ for each $i\in\mathcal F$ gives
\begin{equation}
 \sum_a P_a[T_a]
 =8\sum_{j\notin\mathcal F}[R_j],
 \qquad P_a\in\mathbb Z_{>0}.
 \label{eq:remaining-toric-tadpole}
\end{equation}
For $i\in\mathcal F$, pairing the right-hand side with $[R_i]\wedge J$ gives zero because the O7 components are disjoint. Every remaining class $[T_a]$ differs from $[R_i]$, so its smooth irreducible supports have proper intersections with $R_i$. The left-hand side is therefore a sum of nonnegative intersection integrals, and each integral must vanish. Every remaining D7 support is consequently disjoint from $R_i$:
\begin{align}
    T_a\cap R_i=\varnothing\qquad(i\in\mathcal F).
\end{align}
Here $T_a$ denotes any occupied representative of the class $[T_a]$. The same argument applies to every D7 representative of $[R_i]$: replacing $R_i$ by such a representative leaves the intersection integrals unchanged. Thus the component sectors with $i\in\mathcal F$ are geometrically sequestered from the remaining D7 stacks: every support in these sectors is disjoint from every remaining D7 support. The argument of the proposition then shows that, in every component sector with $i\in\mathcal F$, massless hypercharge is incompatible with nonzero net Standard Model quark-doublet chirality.

Any net chiral Standard Model quark doublets must therefore arise from the remaining stacks, whose charges must satisfy \eqref{eq:remaining-toric-tadpole}. If every O7 component belongs to $\mathcal F$, the right-hand side vanishes. Pairing with $J^2$ forces the remaining sum to be empty, since every occupied support has positive K\"ahler volume. The entire configuration is then componentwise, and the no-go proposition applies.

The allocation test imposes no restriction on the number of smooth irreducible representatives of the allowed classes. If $R_i$ is the unique effective representative of its class for some $i\in\mathcal F$, its eight covering-space D7-branes must lie on the actual component $R_i$. That sector is purely local, with maximal parent algebra $\so(8)$.

A useful application is to favourable Calabi--Yau hypersurfaces obtained from four-dimensional reflexive polytopes in the Kreuzer--Skarke database. Their relevant coordinate divisors are $D_I=X\cap\{x_I=0\}$, with $I=1,\ldots,h^{1,1}(X)+4$~\cite{Batyrev:1993oya,Demirtas:2022hqf}. An inherited toric involution preserving $X$ can give an O7$^{-}$ orientifold with connected O7 components $R_i$. We restrict here to smooth orientifold-compatible Calabi--Yau hypersurfaces, as required by our setup. An O7 component can arise from an ambient fixed divisor intersected with $X$, or from an ambient fixed surface contained in $X$, such as $\widehat D_a\cap\widehat D_b\subset X$~\cite{Moritz:2023jdb}. Its class is orientifold even but need not coincide with an individual coordinate-divisor class $[D_I]$. These constructions provide the divisor-class data needed for the allocation test. The allowed classes may include coordinate-divisor classes and additional classes admitting the required smooth irreducible representatives. Prime toric divisors are used as candidate QCD supports in statistical studies of these geometries~\cite{Gendler:2023kjt,Sheridan:2024vtt,Jain:2025vfh}.

These forced D7 allocations can also affect axion potentials. Statistical axiverse studies often model non-QCD axion potentials using ED3 contributions on prime toric divisors, with suppression $e^{-2\pi\tau_D}$~\cite{Gendler:2023kjt,Sheridan:2024vtt,Jain:2025vfh}. When a component sector with $i\in\mathcal F$ is purely local and its low-energy theory is pure $\mathcal N=1$ $SO(8)$ Yang--Mills, gaugino condensation generates $W_{\mathrm{np},i}\propto e^{-2\pi T_i/6}$, where $T_i=\tau_i+i\theta_i$~\cite{Demirtas:2021nlu}. The weaker exponential suppression can increase the associated axion masses relative to ED3-only estimates. Gauge backgrounds can modify the gauge kinetic function and the light matter content, changing the resulting nonperturbative contributions. We will examine these implications for axion spectra in a separate work~\cite{Jain2026:to_appear_paper1}.

\acknowledgements
This work is supported by the U.S. Department of Energy (DOE) under Grant No. DE-SC0010103. I am thankful to Federico Carta, Michele Cicoli, Jim Halverson, and Jakob Moritz for useful comments and questions.

\appendix

%%%%%%%%%%%%%%%%%%%%%%%%%
\section{Configurations with symplectic factors and massless hypercharge}
\label{app:usp}
%%%%%%%%%%%%%%%%%%%%%%%%%

This appendix gives the detailed derivation of the symplectic argument for pairwise distinct O7 classes presented in Sec.~\ref{subsec:symplectic-proof}. We first determine the allowed Standard Model charges of the eight covering-space Chan--Paton endpoint components in each relevant sector. We then derive the orientifold-projected quark-doublet index and its relation to the flux-induced hypercharge-masslessness condition. The argument covers quark doublets arising within one stack or between different stacks of the same component class.

We assume that the Standard Model generators survive the D7 gauge backgrounds and regard their unbroken algebra as embedded in the parent algebra \eqref{eq:parent}. We allow arbitrary diagonal embeddings:
\begin{equation}
 \begin{gathered}
 \varphi:
 \su(3)_C\oplus\su(2)_L\oplus\uu(1)_Y
 \hookrightarrow\mathfrak g_{\max},\\
 \varphi_i:=\pi_i\circ\varphi,
 \end{gathered}
 \label{eq:SM-projections}
\end{equation}
where $\pi_i$ projects onto the complete algebra $\mathfrak g_i$ of component sector $i$, including all its parent factors. The projections preserve the commutation relations. In particular, the projected color, weak-isospin, and hypercharge subalgebras commute with one another.

\subsection{Color, weak isospin, and hypercharge in one component sector}

Consider a component sector capable of supplying a quark doublet or its conjugate. Both color and weak isospin must act nontrivially on it. Let $V$ be the direct sum of its covering-space Chan--Paton endpoint spaces, including images. It has complex dimension eight, whether the branes occupy one support or several.

The representation on $V$ is equivalent to its complex conjugate. Orthogonal and symplectic endpoint representations have this property individually, while each unitary endpoint representation occurs with its conjugate on the image stack. Consequently, a color representation inequivalent to its conjugate must occur together with that conjugate. A $\mathbf6$ would require an additional $\overline{\mathbf6}$, exceeding eight dimensions. An irreducible color adjoint $\mathbf8$ would use all eight Chan--Paton endpoint components allowed by the componentwise allocation. Schur's lemma would then leave only scalar matrices commuting with color, which cannot realize a nonzero weak $\su(2)$. The only possibility is therefore
\begin{equation}
 V\big|_{\su(3)}
 =\mathbf3\oplus\overline{\mathbf3}
 \oplus\mathbf1\oplus\mathbf1.
 \label{eq:usp-color}
\end{equation}

The color triplet and antitriplet each occur once. Transformations commuting with color can therefore act on each only by an overall phase. Weak isospin must act on the two remaining color-singlet components, which form one doublet.

This doublet must lie in a symplectic parent factor. An orthogonal action on the two remaining components provides only $\so(2)$, which cannot contain $\su(2)$. A doublet on a non-invariant unitary stack would require another doublet on its image, using four color-singlet components rather than the two available.

The possible parent algebras of a component sector supporting both color and weak isospin are therefore:
\begin{equation}
 \begin{gathered}
 \usp(8),\qquad \so(6)\oplus\usp(2),\\
 \uu(3)\oplus\usp(2),\qquad \usp(6)\oplus\usp(2).
 \end{gathered}
 \label{eq:quark-capable-parents}
\end{equation}
The first algebra uses all four brane--image pairs in one stack. The other three use a $3+1$ split. Any multiplicity partition other than $4$ or $3+1$ has $N_{i\alpha}\leq2$ for every stack, and none of its parent factors can contain $\su(3)$. A diagonal embedding cannot evade this: its nonzero projection onto any parent factor would still have to embed $\su(3)$ into that factor. These decompositions agree with the standard subgroup tables \cite{Slansky:1981yr}.

Although the parent algebras in \eqref{eq:quark-capable-parents} differ, their combined endpoint spaces have the same eight-dimensional decomposition under color and weak isospin. We now determine the allowed hypercharge action on these endpoint components. This specifies the charges carried by individual string endpoints; physical D7--D7 fields carry two such indices, and their charges will be obtained below.

A hypercharge generator commuting with color and weak isospin acts by a scalar on each irreducible endpoint block. The color and image-color blocks have opposite eigenvalues. On the weak doublet, an overall phase does not preserve the symplectic pairing, so the commuting hypercharge generator vanishes. The joint decomposition and hypercharge action are therefore
\begin{equation}
 \begin{gathered}
 \left.V\right|_{\varphi_i(\mathfrak g_{\rm SM})} = (\mathbf3,\mathbf1)_{q_i} \oplus(\overline{\mathbf3},\mathbf1)_{-q_i} \oplus(\mathbf1,\mathbf2)_0,
 \end{gathered}
 \label{eq:usp-branching}
\end{equation}
where $\mathfrak g_{\rm SM} =\su(3)_C\oplus\su(2)_L\oplus\mathfrak u(1)_Y$. In this basis, the hypercharge generator is
\begin{equation}
 Y_i=q_i\operatorname{diag}
 \bigl(\mathbf1_3,-\mathbf1_3,\mathbf0_2\bigr).
 \label{eq:usp-Y}
\end{equation}
All eight endpoint components have been used. There is no additional block in this sector that could contribute independently to the hypercharge-masslessness condition.

To obtain the physical matter representations, we must combine the two endpoint indices. A field represented by a map from one endpoint space to another transforms in the target representation and the dual of the source representation. Its hypercharge is therefore the target charge minus the source charge. In particular, a color endpoint of charge $q_i$ and a weak endpoint of charge zero can give a quark doublet of hypercharge $q_i$.

Write the endpoint space as $V=V_c\oplus V_{c'}\oplus V_w$, with dimensions $3$, $3$, and $2$, respectively. The two ordered Chan--Paton blocks carrying quark-doublet charges are
\begin{equation}
 \begin{aligned}
 \Phi &: V_w\longrightarrow V_c,\\
 \widetilde\Phi &: V_{c'}\longrightarrow V_w.
 \end{aligned}
 \label{eq:usp-Q-blocks}
\end{equation}
Thus $\Phi$ is a $3\times2$ matrix and $\widetilde\Phi$ is a $2\times3$ matrix. Under color and weak transformations $U\in SU(3)_C$ and $W\in SU(2)_L$,
\begin{equation}
 \Phi\longmapsto U\Phi W^{-1},
 \qquad
 \widetilde\Phi\longmapsto W\widetilde\Phi U^T.
 \label{eq:usp-Q-transformations}
\end{equation}
The source space $V_{c'}$ transforms by $U^*$, so the source index of a map transforms by the inverse matrix $(U^*)^{-1}=U^T$. This inverse action is what is meant by transforming in the dual of the source. For $SU(2)$, a doublet and its complex conjugate are equivalent representations.\footnote{With $\varepsilon=i\sigma_2$, the identity $\varepsilon W^*\varepsilon^{-1}=W$ explicitly relates the two transformation laws.} Both blocks therefore carry $(\mathbf3,\mathbf2)$.

The corresponding hypercharge actions are
\begin{equation}
 \begin{aligned}
 Y_{c,i}\Phi-\Phi Y_{w,i}&=q_i\Phi,\\
 Y_{w,i}\widetilde\Phi-\widetilde\Phi Y_{c',i}
 &=q_i\widetilde\Phi,
 \end{aligned}
 \label{eq:usp-Q-hypercharge}
\end{equation}
where $Y_{c,i}=q_i\mathbf1_3$, $Y_{c',i}=-q_i\mathbf1_3$, and $Y_{w,i}=\mathbf0_2$. The zero hypercharge of the weak endpoint block therefore does not imply zero hypercharge for a physical weak doublet: the two endpoint charges enter with opposite signs. Both blocks in~\eqref{eq:usp-Q-blocks} carry $(\mathbf3,\mathbf2)_{q_i}$; the reversed maps carry the conjugate representation.

Here $\mathbf3$ is defined relative to the global color algebra in \eqref{eq:SM-projections}. A sector supplying $Q=(\mathbf3,\mathbf2)_{1/6}$ must consequently have $q_i=1/6$. The obstruction below needs only $q_i\ne0$.

\subsection{The quark-doublet index}

Let $C_i$ support the color triplet, let
$C_i'=\sigma(C_i)$ support its image, and let
$W_i=\sigma(W_i)$ support the weak doublet. All three
supports have class $[R_i]$. They may coincide or be distinct;
the following argument covers both possibilities.

A gauge background preserving the projected color and weak
groups must act identically on the components within each
irreducible block of \eqref{eq:usp-branching}. Since each block
occurs once, its internal field strength has the form
\begin{equation}
 \begin{gathered}
 \cF_{c,i}=f_{c,i}\mathbf1_3,\qquad
 \cF_{c',i}=f_{c',i}\mathbf1_3,\\
 \cF_{w,i}=f_{w,i}\mathbf1_2.
 \end{gathered}
 \label{eq:usp-flux}
\end{equation}
The $f$'s denote dimensionless gauge-invariant flux classes,
including the pulled-back $B$ field, on their respective
supports. This form follows from preserving the gauge
generators; it is not an additional choice of Abelian flux
ansatz.

The orientifold exchanges the color and image-color blocks and
preserves the weak block. Its action on the gauge field gives
\cite{Collinucci:2008pf}
\begin{equation}
 f_{c',i}=-\sigma^*f_{c,i},
 \qquad
 f_{w,i}=-\sigma^*f_{w,i}.
 \label{eq:usp-flux-parity}
\end{equation}
The first pullback uses the map $\sigma:C_i'\to C_i$,
and the second uses $\sigma:W_i\to W_i$. Define
\begin{equation}
 \begin{aligned}
 A_{c,i}&:=\int_{C_i}[R_i]\big|_{C_i}\wedge f_{c,i},\\
 A_{c',i}&:=\int_{C_i'}[R_i]\big|_{C_i'}\wedge f_{c',i},\\
 A_{w,i}&:=\int_{W_i}[R_i]\big|_{W_i}\wedge f_{w,i}.
 \end{aligned}
 \label{eq:usp-flux-integrals}
\end{equation}
The notation $[R_i]|_S$ means the bulk divisor class restricted
to the support $S$. Since $[R_i]$ is orientifold-even and
$\sigma$ preserves the complex orientations of these supports,
\eqref{eq:usp-flux-parity} implies
\begin{equation}
 A_{c',i}=-A_{c,i},
 \qquad A_{w,i}=0.
 \label{eq:usp-integral-parity}
\end{equation}
In particular, the integral of the odd weak flux against the
even restricted class vanishes.

The ordinary D7 bifundamental index applies both to intersecting
supports and to coincident gauge blocks; in the latter case it
contains the divisor self-intersection
\cite{Hebecker:2011hk}. For distinct color and weak supports,
the color-flux contribution is
$\int_{C_i}[W_i]|_{C_i}\wedge f_{c,i}=A_{c,i}$, because
$[W_i]=[R_i]$. The weak-flux contribution similarly equals
$A_{w,i}$. The same expressions apply when the supports
coincide.

For an ordered block $a\leftarrow b$, meaning a map $V_b\to V_a$, the source index transforms with the inverse gauge transformation. The two endpoints therefore couple with opposite signs to their gauge backgrounds. Before the orientifold projection, the chiral indices of the blocks in \eqref{eq:usp-Q-blocks} are
\begin{equation}
 \begin{aligned}
 I_{cw}&=\epsilon_i\bigl(A_{c,i}-A_{w,i}\bigr),\\
 I_{wc'}&=\epsilon_i\bigl(A_{w,i}-A_{c',i}\bigr).
 \end{aligned}
 \label{eq:usp-covering-indices}
\end{equation}
Here the indices count copies of the full $(\mathbf3,\mathbf2)$ representation, and $\epsilon_i=\pm1$ sets the common index convention.

The orientifold reverses the endpoint order and replaces each endpoint by its image: $(a\leftarrow b)\mapsto(b'\leftarrow a')$. Since $w'=w$ and $c,c'$ are exchanged, it maps $c\leftarrow w$ to $w\leftarrow c'$ and vice versa. Both blocks carry the same quark-doublet charges.

The projection identifies the modes in these two blocks and keeps one copy of each paired mode. The physical index is therefore half the sum of the two unprojected indices \cite{Collinucci:2008pf}:
\begin{equation}
 \mathcal I_{Q_i}
 =\frac12\bigl(I_{cw}+I_{wc'}\bigr).
 \label{eq:usp-Q-index}
\end{equation}
The two blocks are exchanged rather than individually projected onto symmetric or antisymmetric representations. There is therefore no additional O7 correction for this pair of blocks.

Using \eqref{eq:usp-integral-parity}, both ordinary indices equal $\epsilon_i A_{c,i}$. Consequently,
\begin{equation}
 \mathcal I_{Q_i}=\epsilon_i A_{c,i}.
 \label{eq:usp-Q-even}
\end{equation}
The fluxes need not be restrictions of flux classes defined on the bulk space $X$.

\subsection{The massless-hypercharge condition}

Equation~\eqref{eq:usp-Q-even} expresses the net quark-doublet chirality in terms of a single flux integral, $A_{c,i}$. The same integral determines this sector's flux-induced hypercharge coupling. We will use disjointness to show that a nonzero value necessarily makes hypercharge massive when $q_i\ne0$.

The four-dimensional axions $\theta_\alpha$ arise from the scalar components of the RR four-form:
\begin{equation}
 C_4=\sum_\alpha\theta_\alpha\,\widetilde\omega^\alpha+\cdots,
 \qquad
 \widetilde\omega^\alpha\in H^{2,2}_{+}(X).
 \label{eq:usp-C4-axions}
\end{equation}
Here the $\widetilde\omega^\alpha$ form a basis of the even four-form classes. Let $\omega_\alpha\in H^{1,1}_{+}(X)$ denote the dual basis, normalized by
\begin{equation}
 \int_X\omega_\alpha\wedge\widetilde\omega^\beta
 =\delta_\alpha{}^\beta.
\end{equation}
Dimensional reduction of the D7 Chern--Simons couplings, together with the self-duality of the RR five-form field strength, gauges the shifts of these axions~\cite{Grimm:2011tb,Mayrhofer:2013ara}. The charges are determined by the flux integrals
\begin{equation}
 K_Y(\omega)
 :=
 \sum_S\int_S
 \iota_S^*\omega\wedge
 \operatorname{tr}(Y_S\cF_S),
 \label{eq:usp-K}
\end{equation}
where $S$ denotes an occupied D7 support divisor and $\iota_S:S\hookrightarrow X$ is its inclusion. The sum runs over the distinct supports in the covering space, including orientifold images. All branes sharing the same support are combined into one Chan--Paton bundle on $S$. Here $Y_S$ is the action of the global hypercharge generator on this bundle, and $\cF_S$ is its internal gauge-invariant field strength. The trace includes all Chan--Paton components on $S$.

Under a hypercharge gauge transformation with parameter $\lambda_Y$, the fields transform as
\begin{equation}
 \begin{aligned}
 A_{Y\mu}&\longmapsto A_{Y\mu}+\partial_\mu\lambda_Y,\\
 \theta_\alpha&\longmapsto
 \theta_\alpha+k_\alpha\lambda_Y,
 \qquad k_\alpha\propto K_Y(\omega_\alpha).
 \end{aligned}
 \label{eq:usp-axion-gauging}
\end{equation}
The common orientifold and field normalizations are
absorbed into the proportionality. The gauge-invariant
derivatives and their kinetic terms are therefore
\begin{equation}
 \begin{aligned}
 D_\mu\theta_\alpha
 &=\partial_\mu\theta_\alpha-k_\alpha A_{Y\mu},\\
 \mathcal L_{\rm axion}
 &=-\frac12 G^{\alpha\beta}
 D_\mu\theta_\alpha D^\mu\theta_\beta,
 \end{aligned}
\end{equation}
where $G^{\alpha\beta}$ is the positive-definite axion
kinetic matrix.

Expanding these kinetic terms gives a hypercharge gauge-boson mass coefficient proportional to $G^{\alpha\beta}k_\alpha k_\beta$. This is the St\"uckelberg mechanism. Positivity implies that this coefficient vanishes only when every $k_\alpha$ vanishes. Since the $\omega_\alpha$ form a basis, a necessary condition for massless hypercharge is
\begin{equation}
 K_Y(\omega)=0
 \qquad
 \text{for every }\omega\in H^{1,1}_{+}(X).
 \label{eq:usp-massless-Y}
\end{equation}

This condition applies also when hypercharge is a combination of generators from several sectors: $Y_S$ in \eqref{eq:usp-K} is the projection of that same global generator onto each stack. Other $U(1)$ gauge bosons may be St\"uckelberg massive; the final hypercharge combination must satisfy this condition.

For a component sector with endpoint decomposition \eqref{eq:usp-branching}, let $K_{Y,i}(\omega)$ denote its contribution to \eqref{eq:usp-K}. For an even class $\omega$,
\begin{align}
 K_{Y,i}(\omega)
 &=3q_i\int_{C_i}\omega\big|_{C_i}\wedge f_{c,i} - 3q_i\int_{C_i'}\omega\big|_{C_i'}\wedge f_{c',i}\nonumber\\
 &=6q_i\int_{C_i}\omega\big|_{C_i}\wedge f_{c,i}.
\label{eq:usp-Y-flux-trace}
\end{align}
The last equality follows from the orientifold relation between the color fluxes. The weak block contributes nothing because its hypercharge eigenvalue in \eqref{eq:usp-Y} is zero. When the color and image-color blocks share a support, the two terms are contributions to the trace on that same support; when their supports differ, they are separate terms in \eqref{eq:usp-K}.

\subsection{Disjointness isolates each sector}

Apply \eqref{eq:usp-massless-Y} to the even class $\omega=[R_i]$. Every D7 support $S$ belonging to a different component sector is disjoint from $R_i$, by the geometric argument in the main text. Consequently,
\begin{equation}
 [R_i]\big|_S=0
 \quad\text{in }H^2(S,\mathbb R).
 \label{eq:usp-restriction}
\end{equation}
Indeed, for any closed two-cycle contained in $S$, the integral of $[R_i]|_S$ is its intersection number with $R_i$. This vanishes because $S$ and $R_i$ are disjoint.

All other component sectors therefore drop out of $K_Y([R_i])$. Using \eqref{eq:usp-Y-flux-trace} and
\eqref{eq:usp-Q-even}, we obtain
\begin{equation}
 \begin{aligned}
 0=K_Y([R_i])
 &=K_{Y,i}([R_i])\\
 &=6q_i A_{c,i}
 =6q_i\epsilon_i\,\mathcal I_{Q_i}.
 \end{aligned}
 \label{eq:usp-local-obstruction}
\end{equation}
Since a sector supplying an SM quark doublet has $q_i\ne0$,
\begin{equation}
 \mathcal I_{Q_i}=0.
 \label{eq:usp-zero-index}
\end{equation}
Mixing hypercharge with generators in other component sectors cannot evade this conclusion: their contributions to this test vanish separately. The argument uses disjointness, not linear independence of the component classes.

Finally, consider the arbitrary embedding \eqref{eq:SM-projections}. Equation~\eqref{eq:ext} implies that each massless D7--D7 fermion belongs to one component sector, with Standard Model transformation law determined by $\varphi_i$. Sectors without symplectic factors cannot supply quark doublets by the main-text proof. Every remaining sector that can supply an SM quark doublet has zero net index by
\eqref{eq:usp-zero-index}. Hence
\begin{equation}
 \mathcal I_Q=\sum_i\mathcal I_{Q_i}=0,
\end{equation}
where sectors without the SM quark-doublet representation contribute zero. This proves the proposition for pairwise distinct O7 classes, including split and unsplit allocations, mixed products, and diagonal embeddings. Section~\ref{subsec:repeated-o7-classes} establishes the extension to repeated classes.

The conclusion allows massless vectorlike quark-doublet pairs. It excludes the net chirality required by the Standard Model. The proof uses the fermion index and the D7 Chern--Simons couplings, and does not require the stacks to preserve a common four-dimensional supersymmetry.

\bibliography{bibfile}
\end{document}